\documentclass[10pt,letterpaper,twocolumn]{article} 
\usepackage{ol2}
\usepackage[active]{srcltx}
\usepackage[draft]{hyperref}
\usepackage{amsmath}
\newcommand{\na}{\vec{\nabla}}
\newcommand{\E}{\vec{E}}
\newcommand{\Hb}{\vec{H}}
\newcommand{\ps}{\vec{\psi}}
\newcommand{\n}{\vec{n}}
\newcommand{\rr}{\vec{r}}
\newcommand{\J}{\vec{J}}
\newcommand{\kk}{\vec{k}}
\newcommand\fr{\displaystyle\frac}
\newcommand{\eref}[1]{Eq. (\ref{#1})}
\begin{document}
\twocolumn[
\title{Amplification of light in a glass ball suspended within a gainy medium}

\author{Filipp V. Ignatovich$^{1*}$ and Vladimir K. Ignatovich$^2$}

\address{$^1$Lumetrics Inc, 150 Lucius Gordon Dr, ste 117\\
West Henrietta, N.Y. 14586\\
$^2$Frank Laboratory for Neutron Physics, Joint Institute for Nuclear Research,\\
Dubna, Russia 141980\\
$^*$Corresponding author: ifilipp@gmail.com}

\begin{abstract}
We discuss total internal reflection (TIR) from an interface between glass and active gaseous media and propose an
experiment for strong light amplification.
\end{abstract}

\ocis{140.0140, 140.4480, 140.6810, 240.0240, 240.6645} ]

Light reflection from an interface between two media is determined by the wave equation and the boundary conditions,
which follow from Maxwell's equations. We consider Maxwell's equations in media without free charges, with zero
conductivities $\sigma$ and time-independent permittivities $\epsilon$, $\mu$:
\begin{equation}\label{3a}
-[\na \times \E(\rr,t)] = \mu\frac{\partial}{\partial t}\Hb(\rr,t),
\end{equation}
\begin{equation}\label{4a}
[\na\times \Hb(\rr,t)] = \epsilon\frac{\partial}{\partial t} \E(\rr,t)
\end{equation}
\begin{equation}\label{1aa1}
\na\cdot\epsilon\E(\rr,t) = 0,\qquad \na \cdot \mu\Hb(\rr,t) = 0.
\end{equation}
In a homogeneous medium the parameters $\varepsilon$ and $\mu$ are
constant in space, and (\ref{3a})-(\ref{1aa1}) lead to wave
equations
\begin{equation}\label{e}
\Delta\E(\rr,t)=-\mu\epsilon \fr{\partial^2} {\partial
t^2}\E(\rr,t),\,\, \Delta\Hb(\rr,t)=-\mu\epsilon \fr{\partial^2}
{\partial t^2}\Hb(\rr,t).
\end{equation}

Both equations have plain wave solutions
\begin{equation}\label{pw}
\E(\rr,t)=\E\exp(i\kk\rr-i\omega t),\,\,
\Hb(\rr,t)=\Hb\exp(i\kk\rr-i\omega t),
\end{equation}
and substitution of \eref{pw} into \eref{e} shows that
$k^2=\epsilon\mu\omega^2=\omega^2/c^2$, where
$c=1/\sqrt{\epsilon\mu}$ is the speed of light. However the wave
equations are derived from the Maxwell's equations, and the
outcome of the Maxwell's equations after substitution of the first
\eref{pw} into \eref{3a} or the second \eref{pw} into \eref{4a} is
\begin{equation}\label{eh}
\Hb=\fr1{\mu\omega}[\kk\times\E],\qquad
\E=-\fr1{\epsilon\omega}[\kk\times\Hb].
\end{equation}
Therefore, $\E$ and $\Hb$ are orthogonal to $\kk$ and to each
other, and, if $|\E|=1$, the length of $\Hb$ is
$|\Hb|=\sqrt{\varepsilon/\mu}=1/Z$, where
$Z=\sqrt{\mu/\varepsilon}$ is called the medium impedance.

In the next section we consider the reflection of the light wave
from an interface between two media when the reflecting medium is
lossy or gainy, find peculiarities of the reflection amplitudes,
prove that the TIR reflection coefficient for gainy reflecting
medium is larger than unity and propose an experiment for strong
enhancement of light field.

\section{Wave reflection and refraction at an interface}

If space consists of two halves with different $\epsilon_{1,2}$
and $\mu_{1,2}$, \eref{e} cannot be derived directly from
(\ref{3a})-(\ref{1aa1}), because the permittivities depend on the
coordinates. However, each half is homogeneous and has its own
wave equation with its own plain wave solution. The interface
creates reflection and refraction of the waves, which is
determined by requirements of the Maxwell's equations. According
to them components $\E_\|(\rr,t)$, $\Hb_\|(\rr,t)$ parallel to the
interface, and $\epsilon(\n\cdot\E(\rr,t))$,
$\mu(\n\cdot\Hb(\rr,t))$, perpendicular to it ($\n$ is a unit
normal vector) must be continuous. The wave function in the
presence of the interface becomes
\begin{align}
&\ps(\rr,t)=\nonumber\\
&\Theta(z<0)\Big(\exp(i\kk_1\rr-i\omega t)\ps_1+\exp(i\kk_r\rr-i\omega
t)\ps_r\rho\Big)\nonumber\\
&+\Theta(z>0)\exp(i\kk_2\rr-i\omega t)\ps_2\tau,\label{wf}
\end{align}
where the term $\exp(i\kk_1\rr-i\omega t)\ps_1$ with the wave vector $\kk_1$ describes the plain wave incident on the
interface from medium 1, factors $\ps_i=\E_i+\Hb_i$ ($i=1,r,2$) denote sum of electric and magnetic polarization
vectors, $\kk_{r,2}$ are wave vectors of the reflected and transmitted waves, $\rho$, $\tau$ are the reflection and
transmission amplitudes respectively, and $\Theta(z)$ is the step function, which is equal to unity when inequality in
its argument is satisfied, and to zero otherwise.

The wave vectors $\kk_{r,2}$ are completely determined by $\kk_1$.
They are determined uniquely by the constants $\epsilon_i$,
$\mu_i$, and by the fact that the frequency $\omega$ and the part
$\kk_\|$ of the wave vectors parallel the interface must be
identical for all the waves. In the following we assume that the
medium 1 is lossless, i.e. $\epsilon_1\mu_1$ is real, therefore
all the of the components of $\kk_1$ are also real.

The normal component $\kk_{2\bot}$ of the refracted wave can be
represented as
\begin{equation}
k_{2\bot}=\sqrt{\epsilon_2\mu_2 k_1^2-\kk_\|^2} =\sqrt{k_{1\bot}^2
-(\epsilon_1\mu_1-\epsilon_2\mu_2)k_1^2}, \label{ep2n}
\end{equation}
or
\begin{equation}\label{ep2}
k_{2\bot}=\sqrt{\epsilon k_1^2-\kk_\|^2}=\sqrt{n^2
k_1^2-\kk_\|^2},
\end{equation}
where $n=\sqrt{\epsilon}$ is the refractive index, and we
introduced relative permittivity
$\epsilon=\epsilon_2\mu_2/\epsilon_1\mu_1$.

From \eref{ep2} it follows that for lossless media when
$0<\epsilon<1$ is real, the incident wave, for which $\kk_\|$ is
within $nk_1\le |\kk_\||\le k_1$, is totally reflected from the
interface. This happens because
\begin{equation}\label{ep3}
k_{2\bot}=iK''_{2\bot}\equiv i\sqrt{k_\|^2-\epsilon k_1^2},
\end{equation}
thus the factor $\exp(ik_{2\bot}z)=\exp(-K''_{2\bot}z)$ of the
wave $\exp(i\kk_2\rr)$ exponentially decays, and the refracted
wave becomes an evanescent one. Therefore, the energy does not
flow inside the medium 2, and due to the energy conservation it
must be totally reflected into medium 1.

If the medium 2 is lossy or gainy, the constant $\epsilon$ is a
complex quantity $\epsilon=\epsilon'\pm i\epsilon''$, with
positive $\epsilon'$ and $\epsilon''$. In this case outside the
total internal reflection (TIR) region ($|\kk_\||^2\ll
\epsilon'k_1^2$) we have $k_{2\bot}= k'_{2\bot}\pm ik''_{2\bot}$,
where for small $\epsilon''$ ($\epsilon''k_1^2\ll
\epsilon'k_1^2-|\kk_\||^2$)
\begin{equation}\label{eps5}
k'_{2\bot}\approx\sqrt{\epsilon'k_1^2-|\kk_\||^2},\quad
k''_{2\bot}\approx\epsilon''\fr{k_1^2}{2k'_{2\bot}}.
\end{equation}

In the TIR regime, $k'_{2\bot}$ in Eq. \ref{eps5} transforms into
$iK''_{2\bot}$, where
$K''_{2\bot}\approx\sqrt{|\kk_\||^2-\epsilon'k_1^2}$,  and
$k''_{2\bot}$ transforms to
\begin{equation}\label{eps6}
k''_{2\bot}\to -iK'_{2\bot}=\epsilon''\fr{k_1^2}{2iK''_{2\bot}}.
\end{equation}
Therefore, in TIR $k_{2\bot}=\pm K'_{2\bot}+iK''_{2\bot}$, where
\begin{equation}\label{ep8}
K'_{2\bot}=\epsilon''\fr{k_1^2}{2K''_{2\bot}},\qquad K''_{2\bot}\approx\sqrt{|\kk_\||^2-\epsilon'k_1^2}.
\end{equation}
The '$+$' sign before imaginary part $iK''_{2\bot}$ determines
exponential decay of the refracted wave away from the interface
for both lossy and gainy media cases. However the real part,
$K'_{2\bot}$ has opposite signs for lossy and gainy cases. The
positive value of $K'_{2\bot}$ for lossy medium means that the
reflection coefficient in TIR is less than one, because part of
the energy flux proportional to $K'_{2\bot}$ enters the medium 2
and decays there. The negative value of $K'_{2\bot}$ for gainy
medium means that the reflection coefficient in TIR is larger than
one, because part of the energy flux proportional to $K'_{2\bot}$,
exits the medium 2 and adds to the TIR wave.

By the way, we want to note here that the widely spread belief
that the energy flux is given by the Pointing vector
$\J=[\E\times\Hb]$ is not correct. The energy flux is given by
\begin{equation}\label{en}
\J=c\fr{\kk}{k}\fr{\epsilon E^2+\mu H^2}{8\pi},
\end{equation}
i.e. it is equal to the energy density times the light speed, and
it has direction along the wave vector $\kk$. For a plain wave
this definition coincides with the Pointing vector. However the
latter can be defined for wider varieties of vectors $\E$ and
$\Hb$, including stationary fields and evanescent waves for which
the Pointing vector has no relation to the energy flux.

\subsection{Reflection and refraction amplitudes}

The procedure for calculating the reflection amplitude in general case is well explained in~\cite{land}, so here we
only briefly recall it. The polarization $\E_1$ of the incident wave can be arbitrary (except it must be perpendicular
to the wave-vector $\kk_1$). It can be decomposed as $\E=\E_{1s}+\E_{1p}$, where $\E_{1s}$ is the component parallel to
the interface and perpendicular to the plane of incidence (plane of vectors $\kk_1$ and the normal $\n$ to the
interface), and where $\E_{1p}$ lies in the plane of incidence. Reflection amplitudes for each component are different
and can be found independently.

Lets find the reflection of the wave $\E_{1s}$ (s-wave or TE-wave). The field $\E_{1s}$ is accompanied by the field
$\Hb_{1p}$, which lies in the plane of incidence. The total wave function of the TE- wave according to Eq. \ref{wf} can
be represented as $\exp(i\kk_\|\rr_\|-i\omega t)\ps(z)$, where
\begin{align}
\ps(z)&=\Theta(z<0)\Big[\ps_{1s}\exp(ik_{1\bot}z)+ \ps_{rs}\rho_s\exp(-ik_{1\bot}z)\Big]\nonumber\\
&+ \Theta(z>0)\ps_{2s}\tau_s\exp(ik_{2\bot}z), \label{pw4a}
\end{align}
and for $i=1,r,2$ we introduced notations
\begin{equation}\label{pw4b}
\ps_{is}=\E_{1s}+\Hb_{ip},\,\,\,
\Hb_{ip}=\fr{1}{\mu_i\omega}[\kk_i\times\E_{1s}],\,\,\,
\mu_r=\mu_1.
\end{equation}
The corresponding wave vectors are
\begin{equation}\label{pw4c}
\kk_1=\kk_\|+\n k_{1\bot},\,\,\, \kk_r=\kk_\|-\n k_{1\bot},\,\,\,
\kk_2=\kk_\|+\n k_{2\bot}.
\end{equation}

Maxwell's equations require continuity of the electric field $\E_{s}$ at the interface, which leads to the equation
$1+\rho_s=\tau_s$. The same requirement for the component $\Hb_{\|p}$ of the magnetic field parallel to the interface
leads to the equation $(1-\rho_s)k_{\bot1}/\mu_1=\tau_sk_{\bot2}/\mu_2$. The third requirement for the continuity of
the quantity $\mu(\n\cdot\Hb_{p})$ leads to the same equation $1+\rho_s=\tau_s$ as the one obtained from the continuity
of $\E_{s}$. Therefore we have two independent equations, from which we obtain the well known Fresnel formulas
\begin{equation}\label{s1}
\rho_s=\fr{\mu_2k_{1\bot}-\mu_1k_{2\bot}}{\mu_2k_{1\bot}+\mu_1k_{2\bot}},\qquad
\tau_s=\fr{2\mu_2k_{1\bot}}{\mu_2k_{1\bot}+\mu_1k_{2\bot}}.
\end{equation}

Similar considerations of the TH-wave with $\E_{1p}$ polarization gives the other two expressions,
\begin{equation}\label{s2}
\rho_p=\fr{\epsilon_2k_{1\bot}-\epsilon_1k_{2\bot}}{\epsilon_2k_{1\bot}+\epsilon_1k_{2\bot}},\qquad
\tau_p=\fr{2\epsilon_2k_{1\bot}}{\epsilon_2k_{1\bot}+\epsilon_1k_{2\bot}}.
\end{equation}
For simplicity, we limit ourselves only to TE-case and assume that
$\mu_2=\mu_1$, so $\mu_{1,2}$ in Eqs. (\ref{s1}) is excluded.

Once can see that in TIR the reflection coefficient from a gainy
medium is larger than one, and it increases with gain. The growth
of the reflection coefficient can be explained by the photon
emission toward the interface, induced by the evanescent field.
The increase in the reflected flux is due to the sub-barrier
induction of the photon, which tunnels from the gainy medium into
medium 1 and coherently adds to the reflected primary photon. The
larger is the gain, the larger is the probability of such process.

\section{The experiment for strong enhancement of the light trapped in a
glass sphere}

The increase of the reflection coefficient at TIR from a gainy
medium can be used to develop a curious experiment for storage and
amplification of light. Imagine a glass sphere with a coupler $P$,
as shown in Fig.1. The sphere has thin walls (it is also possible
to use a homogeneous glass sphere) and is surrounded by an excited
gas (or other active media). The ray of light, shown by the thin
line, enters the glass walls through the coupler and then
undergoes TIR multiple times. At every reflection the light is
amplified according to the analysis in the previous Section. At
the end the ray escapes the sphere, as shown by the thick line.
The amplification depends on the number of the reflections and on
the gain coefficient of the active medium. The number of the
reflections is very sensitive to the angle of the incident ray. If
the overall amplification is sufficiently high, the glass will
melt into a liquid bubble with thin skin filled with the light,
similar to the ball lightning~\cite{ig}.

\begin{figure}[h!]
\centering\includegraphics[width=1.5in]{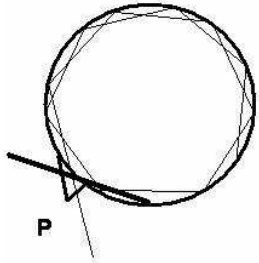}
\caption{\label{f2} Schematic of the experiment for multiple TIR
off gainy medium.}
\end{figure}

We can estimate the magnitude of the light enhancement in such a
sphere. Assume that for the active medium
$\epsilon_2\approx1-i\alpha$, and $\epsilon_1$ of the glass is
real. For TEM mode, the reflection amplitude at TIR according to
Eq. \ref{s1} can be written as
\begin{align}
\rho_s&=\fr{k_{1\bot}-i\sqrt{(\epsilon_1-1+i\alpha)k^2-k^2_{1\bot}}}
{k_{1\bot}+i\sqrt{(\epsilon_1-1+i\alpha)k^2-k^2_{1\bot}}}\nonumber\\
&\approx\fr{k_{1\bot}-iK_{2\bot}+\alpha k^2/2K_{2\bot}} {k_{1\bot}+iK_{2\bot}-\alpha k^2/2K_{2\bot}},\label{b1}
\end{align}
where $K_{2\bot}=\sqrt{(\epsilon_1-1)k^2-k^2_{1\bot}}$, and the approximation is valid for small $\alpha$. From Eq.
\ref{b1} follows that the reflection coefficient is
\begin{equation}
|\rho_s|^2=\fr{[k_{1\bot}+\alpha
k^2/2K_{2\bot}]^2+K^2_{2\bot}}{[k_{1\bot}-\alpha
k^2/2K_{2\bot}]^2+K^2_{2\bot}}\approx1+2\alpha
\fr{k_{1\bot}}{K_{2\bot}(\epsilon_1-1)}.\label{b2}
\end{equation}
For the TH mode,
\begin{align}
\rho_p&=\fr{\epsilon_2k_{1\bot}-\epsilon_1k_{2\bot}}{\epsilon_2k_{1\bot}+\epsilon_1k_{2\bot}}
\nonumber\\
&\approx\fr{(1-i\alpha)k_{1\bot}-i\epsilon_1K_{2\bot}+\alpha \epsilon_1k^2/2K_{2\bot}}
{(1-i\alpha)k_{1\bot}+i\epsilon_1K_{2\bot}-\alpha \epsilon_1k^2/2K_{2\bot}},\label{b3}
\end{align}
and
\begin{align}
|\rho_p|^2&=\fr{[k_{1\bot}+\alpha\epsilon_1 k^2/2K_{2\bot}]^2+(\alpha
k_{1\bot}+\epsilon_1K_{2\bot})^2}{[k_{1\bot}-\alpha\epsilon_1 k^2/2K_{2\bot}]^2+(\alpha
k_{1\bot}-\epsilon_1K_{2\bot})^2}
\nonumber\\
&\approx1+2\alpha\epsilon_1\fr{k_{1\bot}}{K_{2\bot}}\fr{k^2+2K^2_{2\bot}}{k^2_{1\bot}+\epsilon^2_1K_{2\bot}^2},\label{b4}
\end{align}
For estimating purposes we can assume that for both cases the
light is amplified by approximately $1+2\alpha$ after each
reflection. Therefore enhancement of the light intensity I after
$N$ collisions with the wall is $I/I_0=(1+2\alpha)^N=\exp(2\alpha
N)$, where $I_0$ is initial intensity. The number of collisions
can be represented as $N=t/t_1$, where $t_1$ is the time between
two consecutive collisions. In sphere of radius $R$ it is
$t_1=2R\sqrt\epsilon\sin\theta/c$, where $\theta$ is the grazing
angle at the collision point. With all these considerations we can
write
\begin{equation}\label{enh}
I/I_0=\exp(t/\tau)\quad{\rm where}\quad
\tau=R\sqrt\epsilon\sin\theta/c\alpha.
\end{equation}

The following analysis is used to estimate $\alpha$. The amplification of a laser beam along a path $l$ inside a gainy
media is $\exp(2k''l)$, where $k''$ is the imaginary part of the wave number, and $g=2k''$ is called the gain
coefficient. In a medium with $\epsilon=1-i\alpha$, the gain coefficient is $g\approx \alpha k=2\pi\alpha/\lambda$,
where $\lambda$ is the wavelength. For N$_2$,CO$_2$ gas lasers, the gain coefficient is approximately $10^{-2}$
cm$^{-1}$~\cite{gas}. For $\lambda/2\pi\simeq10^{-4}$ cm we obtain $\alpha=10^{-6}$. Therefore, the number of the
reflections off the interface should be larger than $N=10^6$ to obtain any practical light amplification.

In the past, many experiments were performed with the whispering
gallery mode resonators (WGMR) of small dimensions ($\sim1$ mm)
and large Q-factors (up to $Q\sim 10^{10}$)~\cite{q}, where light
undergoes large number $N\sim Q$ total internal reflections.

Let's imagine a sphere of $R=10$ cm submerged into an active
medium with $\alpha\sim10^{-7}$, and WGM with $\theta=0.1$, then,
according to \eref{enh}, $1/\tau\approx3\cdot10^3$. Therefore if
initial energy $I_0=10^{-19}$ J, then after 20 ms the energy
inside the resonator will increase enormously up to 10 MJ. The
stored photons will heat and melt the resonator, but the
electrostriction forces will hold the melted substance together.
One can expect to see many interesting nonlinear phenomena in such
systems.

\section*{Acknowledgement}

We express our gratitude to A.Siegman, whose paper~\cite{sig} (see also~\cite{sig1}) inspired us for this work.

\section{History of submissions and rejections~\cite{arx}}
This paper appeared after we tried to criticize the
work~\cite{sig} and sent the critical paper to OPN. The main point
of~\cite{sig} was that reflectivity from a gainy medium at TIR is
smaller than unity like reflectivity from a lossy medium. This
conclusion was deduced from incorrect choice of the sign of the
square root of the complex number
$Z=k^2_{1\bot}-(\epsilon_1-1+i\alpha)k^2$. Dr. A.Siegmann chose
$\sqrt{Z}=-i\sqrt{(\epsilon_1-1+i\alpha)k^2-k^2_{1\bot}}$, while
the correct sign is
$\sqrt{Z}=i\sqrt{(\epsilon_1-1+i\alpha)k^2-k^2_{1\bot}}$, as is
used in \eref{b1}. The choice of sign is related to physics. With
correct sign the wave function inside reflecting medium becomes
$\propto\exp(-K_{2\bot}z-ig_2z)$, where
$K_{2\bot}=\sqrt{(\epsilon_1-1)k^2-k^2_{1\bot}}$ and $g_2=\alpha
k^2/2K_{2\bot}$. This wave function exponentially decays and wave
number $g_2$ gives a flux from reflecting gainy medium, which
makes reflectivity larger than unity. With incorrect sign the wave
function inside reflecting medium becomes
$\propto\exp(K_{2\bot}z+ig_2z)$. For Dr. A.Siegman it means that
because of $g_2$ there is a flux toward the reflecting medium, and
the gain leads to exponential grows of the field. However he did
not notice that the exponential grows does not depend on gain
$\alpha$, and since $K_2\sim1/\lambda\sim10^{-4}$ \AA$^{-1}$, then
at distance $z=1$ mm from the reflecting interface the field
intensity becomes $\exp(2K_{2\bot}z)\sim\exp(2\cdot10^3)\gg
10^{860}$ (i.e. astronomically) higher than the field in the
incident light. So there is a terrible violation of the energy
conservation.

We tried to publish a paper in OPN with careful analysis
of~\cite{sig}, but we were permitted to publish\cite{sig1} only a
list of incorrect claims in~\cite{sig}. However during preparation
of~\cite{sig1} bearing in mind the model of the ball
lightning~\cite{ig} we were struck by an idea that enhancement of
reflectivity from a gainy media at TIR can be used for high
accumulation of energy in a spherical shell and for checking the
idea~\cite{ig} if to put the spherical shell in a gainy medium and
to send a light beam in it in a whispering gallery mode (WGM). So
we submitted this paper to Optics Letters in beginning of June.

On 23 June we received a rejection from Topical Editor of Optics
Letters Timothy Carrig with two referee reports.

\subsection{Referee 1}
I'm sending you herewith my comments on the Ignatovich ms.

Because this is a situation where the Ignatovichs and I have a
fundamental technical disagreement, I do not believe I should
submit this as a formal review, or at least I should not express
any opinion on whether this ms should or should not be published,
leaving that to your editorial judgement and the opinions of other
independent referees.

Rather I am simply expressing my views on the technical matter
itself, for whatever use these may be to you.  In addition, I do
not think I should do this anonymously, and I am therefore copying
this response to the Ignatovichs also.

The major portion of this ms (19 of the 23 equations) is devoted
to a straightforward analysis of the Fresnel reflection
coefficient at a dielectric interface when the lower-index
reflecting medium may have either loss or gain.  Although I have
not checked every step of the derivation, this is in essence a
"textbook problem" and I have no reason to believe that their
analytical results are not correct.

There is, however, a major disagreement between us at one step in
the analysis.  At a point in the general region of Eqs. (11)
through (13) in the ms, one must take the square root of a certain
quantity which becomes complex-valued for either a lossy or gainy
reflecting medium; and one must make a choice of which sign one
chooses for this square root.

For the case of a lossless or lossy medium, the conventional
choice of sign is the one which makes the fields decay away
exponentially with distance into the reflecting medium, a
condition commonly referred as making the fields in this region
evanescent.  The authors of this ms believe that one should make
the same choice for the case of a gainy medium, while I am
convinced that the opposite choice is the physically required or
physically meaningful solution in this case.

My belief is that the real physical constraint on the solution in
both lossy and gainy cases is that the solution to be chosen
should correspond to energy or signals which have arrived from the
higher-index medium travelling on outward at a small but finite
angle (as the equations indicate) into the lower-index medium,
being attenuated or amplified as they go.  I would note that this
is in fact the condition that is met by the "evanescent" lossy
solution, where the direction of energy flow is slightly outward
away from the interface, with the fields decaying in amplitude
with distance as they travel outward because they are travelling
in a lossy rather than a gainy medium.  The same condition should
also govern in the gainy case.

This solution seems to me to be mandated by simple causality, as
well as giving physically reasonable answers for many other
specific waveguide and other specific situations I have examined.

Whether this opinion be right or wrong, however, I must also note
that what I believe to be exactly the same analysis and discussion
of Fresnel reflectivity from a gainy or lossy medium as in this ms
has already been presented, and the same conclusions as in this ms
have been argued, in an OSA journal publication some seven years
ago, i.e.

J. Fan, A. Dogariu, and L.-J. Wang, "Amplified total internal
reflection," Opt. Expr., vol. 11, pp. 299--308, (2003).

Abstract:  Totally internal reflected beams can be amplified if
the lower index medium has gain. We analyze the reflection and
refraction of light, and analytically derive the expression for
the Goos-H\"anchen shifts of a Gaussian beam incident on a
lower-index medium, both  active and absorptive. We examine the
energy flow and the Goos-H\"anchen shifts for various cases. The
analytical results are consistent with the numerical results. For
the TE  mode, the Goos-H\"anchen shift for the transmitted beam is
exactly half of that of the reflected beam, resulting in a "1/2"
rule.

As always, slightly different notation and terminology has been
used,  but the basic problem of Fresnel reflection and its
analysis is straightforward, and I believe this earlier
publication and the authors' current ms present essentially
identical results.  Unless the present authors can point to
genuine differences between this analysis and theirs, I believe
the content and conclusions of this earlier publication are
essentially the same as theirs.  (This ms is also referenced in my
recent OPN article, with an implicit indication that I disagree in
exactly similar fashion with its conclusions.)

The current authors also make brief mention of a proposed
experimental test involving a sphere (or cylinder?) embedded in a
dye medium.  On this I would comment that a planar waveguide or
cylindrical fiber would seem to me a simpler structure for such a
test; that a finite-thickness gain layer is in fact very different
from an unbounded gain layer for such a test in that it has two
reflecting surfaces and will thus display regenerative effects;
and in any case these ideas have already been discussed and even
tested in several publications listed below.

Yours truly,  Tony Siegman

REFERENCES

N. Periasamy and Z. Bor, "Distributed feedback laser action in an
optical fiber by evanescent field coupling," Opt. Commun., vol.
39, pp. 298-302, (1981).

N. Periasamy, "Evanescent wave-coupled dye laser emission in
optical fibers," Appl. Opt., vol. 21, p. 2693, (August 1982).

G. J. Pendock, H. S. Mackenzie, and F. P. Payne, "Dye-Lasers Using
Tapered Optical Fibers," Appl. Opt., vol. 32, pp. 5236--5242,
(1993).

H. Fujiwara and K. Sasaki, "Lasing of a microsphere in dye
solution," Japan. J. Appl. Phys., vol. 38, pp. 5101--5104, (1999).

H. J. Moon, Y. T. Chough, and K. An, "Cylindrical microcavity
laser based on the evanescent-wave-coupled gain," Phys. Rev.
Lett., vol. 85, pp. 3161--3164, (2000).

Y. S. Choi, H. J. Moon, K. Y. An, S. B. Lee, J. H. Lee, and J. S.
Chang, "Ultrahigh-Q microsphere dye laser based on evanescent-wave
coupling," J. Korean Phys. Soc., vol. 39, pp. 928--931, (2001).

{\bf Our comment}

On one side it is respectable that Dr. A.Siegman disclosed himself
as a referee. On the other side it is seen that his main goal was
to prevent publication of our paper. To do that he used the
following arguments: 1) The major portion of manuscript contains
nothing new comparing to textbooks, and I do not agree with
authors (therefore he doesn't agree with textbooks). 2) The
proposed experiment is not interesting. It is better to check with
plain wave guides or fibers. And if to speak about sphere than
there is also nothing new, because there are already so many
publications on micro spheres.

\subsection{Referee 2}

From the report it is seen that the rejection is the main goal of
the referee It is evident from the choice of words. We emphasized
them in bold face without comments.

I am rejecting this paper based on my comments below.

The paper consists of two sections. Section 1 is a  {\bf poor
treatment} of electromagnetic wave reflection and and refraction
at an interface. {\bf A better treatment} can be found in any
standard optic textbook (e.g., Hecht). {\bf Unfortunately}, it has
little bearing on solving for the mode structure of a dielectric
sphere. Solving for the mode structure of the resonances of a
dielectric sphere in vacuum is a classic problem in electricity
and magnetism, and the resulting field distributions have been
known for some time (e.g., Stratton: Electricity and Magnetism.)

{\bf There are many mistakes and misleading assumptions} used
throughout the paper. For example, in section 1 the authors state
that "Maxwell's equations require continuity of the electric field
$E_s$ at the interface..."  However, while the tangential
component of the electric field must be continuous at the surface
to satisfy the boundary conditions, there is a discontinuity in
the radial component of the electric field at the dielectric
boundary. The authors state that they are limiting themselves to
the TE case, however, the TM modes are the interesting ones for
the Whispering Gallery Modes (WGM) of microspheres, since their
electric field vectors are predominantly radial.

Section 2 uses the treatment from Section 1 and applies it to the
WGMs of a dielectric sphere {\bf without using the proper
treatment} for a sphere where the mode structure would be based on
spherical Hankel functions given the boundary conditions. There
are several treatments of this in the literature already (e.g.,
PHYSICAL REVIEW A 67, 2003 033806). It is crucial that the proper
mode structure be accounted for with the proper boundary
condition, especially if it is to be applied in a case where gain
will be present. In that case, there will also be time
dependencies that may need to be addressed for a proper treatment.

{\bf The only new items in the entire paper} consist of {\bf a few
sentences of conjecture} on the amplification of light using the
WGMs of a dielectric sphere immersed in a gain medium.  This
conjecture {\bf is not based on a proper treatment} of the
electromagnetic modes, and {\bf there is no testable prediction}
of the effect.  The {\bf paper amounts to a comment on an idea and
needs serious expansion with a far more careful treatment} before
it could possibly be considered for any peer reviewed journal.

\subsection{Our appeal}

Dear Dr. Carrig,

We would like to appeal your decision due to the reasons listed
below.

The Referee 1, Dr. A.Siegman, is not suitable for reviewing our
paper. As indicated by our earlier email (included at the end of
this email) to you as well as by {\bf Dr. A.Siegman himself}, his
review represents a conflict of interests. We wrote our paper
after reading Dr. Siegman's article in OPN, and we think his views
are erroneous down to the textbook optics level. Obviously, he
does not agree with our point of view and therefore is not
interested in any publications that would criticize his article.
Therefore, we think that Dr. Siegman's comments should not be
counted toward the publication decision, and the manuscript should
be resubmitted to another referee.

We disagree with the decision of Referee 2. He completely
misunderstood our article. We prove this point by addressing the
Referee's review as follows.

1) Section 1 is a  {\it poor treatment of electromagnetic wave
reflection.}

It is unclear what the Referee 2 means by "poor treatment". Is it
a wrong treatment? Absolutely not, it is in fact a textbook
treatment of the electromagnetic wave reflection. Does the Referee
2 thinks that textbooks are poor? Such emotional statement does
not carry any substantiality or significance.

2) {\it Unfortunately, it has little bearing on solving for the
mode structure of a dielectric sphere.}

Judging by the other parts of the review, the Referee's expertise
lays in microspheres. It indeed would be important to reference
the electromagnetic distribution and its effects in a microsphere
if that was the case. However, in our manuscript, the radius of
the sphere is supposed to be much larger than the light wave
length, therefore every reflection can be treated in plane
geometry approximation. The Referee appears to have overlooked
this important detail.

3) {\it There are many mistakes and misleading assumptions}

While this statement would be very important in making publication
decision, {\bf it is not backed by any proof}. The Referee 2 {\bf
does not} point out where in the manuscript he saw the mistakes
and misleading assumptions. His only hint at a mistake is wrong -
see the next point.

4){\it Authors state that "Maxwell's equations require continuity
of the electric field $E_s$ at the interface..."  However, while
the tangential component of the electric field must be continuous
at the surface to satisfy the boundary conditions, there is a
discontinuity in the radial component}

The Referee 2 is correct - the normal component has discontinuity.
However we are talking about the $E_s$ which is the {\bf
tangential} field and {\bf does not} contain the normal component.
Referee 2 overlooked this important detail and made a completely
correct statement appear as incorrect.

5) {\it The authors state that they are limiting themselves to the
TE case, however, the TM modes are the interesting ones for the
Whispering Gallery Modes (WGM) of microspheres, since their
electric field vectors are predominantly radial.}

See the point 2 (we do not consider microspheres). Also, it is
widely accepted in science to consider only one case, when other
cases have similar outcomes for simplicity, clarity and
space-saving purposes, especially when the manuscript pages are
limited. We are not sure why the Referee 2 holds us at fault for
it.

6) {\it Section 2 uses the treatment from Section 1 and applies it
to the WGMs of a dielectric sphere without using the proper
treatment for a sphere where the mode structure would be based on
spherical Hankel functions given the boundary conditions. There
are several treatments of this in the literature already (e.g.,
PHYSICAL REVIEW A 67, 2003 033806). It is crucial that the proper
mode structure be accounted for with the proper boundary
condition, especially if it is to be applied in a case where gain
will be present.}

Again, see the point 2 - we do not consider microspheres.

6) {\it In that case, there will also be time dependencies that
may need to be addressed for a proper treatment.}

The time dependence is a primitive exponential (harmonic). We do
not anticipate any other time dependencies in our configuration.
Again, the Referee thinks of microspheres.

7) {\it The only new items in the entire paper consist of a few
sentences of conjecture on the amplification of light using the
WGMs of a dielectric sphere immersed in a gain medium.}

The use of the word "only" in regards to new items is an emotional
belittling of our ideas. We regard this comment as Referee's
acknowledgement of the novelty of the article, but his view of its
general impact as subjective. These ideas may not epitomize a
breakthrough in science, however they provides direction for
further scientific research, just like the majority of peer-review
articles do.

9) {\it there is no testable prediction of the effect.}

While this particular Referee does not have an idea for an
experimental testing of the effects that we predict, it does not
mean that the ideas are not worth publishing. First, with such
attitude, for example, none of the string theory articles would
ever be published. Second, there are many other scientists who
read OL and it is our hope and goal that someone will become
interested and will have an idea to attempt the experimentally
implementation of the proposed configuration.

The Referee 2 hasn't found any mistakes in our article. He even
acknowledges that our article has new material, however
immediately belittles it. This new material is the main idea
behind our paper, but the referee fails to recognize it instead of
concentrating on other parts of the manuscript in his criticism
from the erroneous microsphere standpoint. We think that the
review by Referee 2 also cannot be used.

We would like to modify the article to emphasize important points
to avoid misunderstandings similar to Reviewer 2 and resubmit the
manuscript.

{\bf The letter we sent you in response to Dr. Siegman's email:}

Dear Dr. Carrig,

We echo Dr. Siegman's concern that it may be advantageous to
forward our manuscript to a third party since our manuscript
originally stems from the fundamental disagreement with Dr.
Siegman's views. We are sorry that we have not made it clear
during submission that Dr. Siegman may not be an appropriate
choice for a reviewer. We appreciate that Dr. Siegman recognized
this conflict of interests and included us in his reply.

We were taken aback when we saw Dr. Siegman's paper in OPN, and we
wanted to remind the scientific community how to calculate Fresnel
reflection amplitudes, which is discussed in many text books and
additionally in the paper Opt. Expr., vol. 11, pp. 299--308,
(2003) cited here by Dr. A.Siegman. We also discussed why the sign
of square root is uniquely defined by physical reasons. If Dr.
A.Siegman were right, electric field in the gainy medium just 1 cm
away from interface would be 4 orders of magnitude larger ({\bf we
mistaken here, look above: it will be more than 800 orders of
magnitude}) than in the total internal reflection field,
notwithstanding how small is the gain.

Our main goal for submitting the manuscript was to propose and
motivate the laser physicists to make an experiment with a
spherical beam trap where the light beam is strongly amplified
after many total internal reflections. Such amplification then
melts the sphere and transforms it to a liquid drop, with physical
processes inside becoming very similar to a ball lightning. Planar
wave guides or cylindrical fibers are unfortunately not
appropriate for this goal.

\subsection{Response from the editor}

After reading your appeal and taking it into consideration, I have
decided to stand by my original decision. Despite your a priori
disagreement with Prof. Siegman, who was cited in your references,
he is well-respected in the field and is a reasonable choice as a
reviewer.  The paper was also sent to a second independent
reviewer.  This reviewer is also a scientist with a very strong
reputation.  This reviewer also recommended rejection, albeit for
somewhat different reasons.  I consider these two reviews adequate
for adjudicating the manuscript.

\subsection{Reply to the editor}

Dear Dr. Carrig,

Is there a next step in the appeal process, where your decision
can be reviewed by another editor? There has to be a reason why in
the judicial system appeals go to an appellate court, different
from the court that decided the original verdict.

It is very unfortunate that modern science puts reputation above
reason. As a recent graduate I do not stand a chance in any
debate, however solid my arguments may be. Such approach in no way
stimulates progress.

Sincerely, Filipp Ignatovich



\end{document}